\documentstyle[psfig]{mn}
\newif\ifAMStwofonts

\ifoldfss
  \ifCUPmtlplainloaded \else
    \NewTextAlphabet{textbfit} {cmbxti10} {}
    \NewTextAlphabet{textbfss} {cmssbx10} {}
    \NewMathAlphabet{mathbfit} {cmbxti10} {} 
    \NewMathAlphabet{mathbfss} {cmssbx10} {} 
  \fi
  \ifAMStwofonts
    \ifCUPmtlplainloaded \else
      \NewSymbolFont{upmath} {eurm10}
      \NewSymbolFont{AMSa} {msam10}
      \NewMathSymbol{\upi}     {0}{upmath}{19}
      \NewMathSymbol{\umu}     {0}{upmath}{16}
      \NewMathSymbol{\upartial}{0}{upmath}{40}
      \NewMathSymbol{\leqslant}{3}{AMSa}{36}
      \NewMathSymbol{\geqslant}{3}{AMSa}{3E}

    \fi
  \fi
\fi 

\ifnfssone
  \newmathalphabet{\mathit}
  \addtoversion{normal}{\mathit}{cmr}{m}{it}
  \addtoversion{bold}{\mathit}{cmr}{bx}{it}
  \newmathalphabet{\mathbfit} 
  \addtoversion{normal}{\mathbfit}{cmr}{bx}{it}
  \addtoversion{bold}{\mathbfit}{cmr}{bx}{it}
  \newmathalphabet{\mathbfss} 
  \addtoversion{normal}{\mathbfss}{cmss}{bx}{n}
  \addtoversion{bold}{\mathbfss}{cmss}{bx}{n}
  \ifAMStwofonts
    \ifCUPmtlplainloaded \else
      \UseAMStwoboldmath
      \makeatletter
      \new@mathgroup\upmath@group
      \define@mathgroup\mv@normal\upmath@group{eur}{m}{n}
      \define@mathgroup\mv@bold\upmath@group{eur}{b}{n}
      \edef\UPM{\hexnumber\upmath@group}
      \new@mathgroup\amsa@group
      \define@mathgroup\mv@normal\amsa@group{msa}{m}{n}
      \define@mathgroup\mv@bold\amsa@group{msa}{m}{n}
      \edef\AMSa{\hexnumber\amsa@group}
      \makeatother
      \mathchardef\upi="0\UPM19
      \mathchardef\umu="0\UPM16
      \mathchardef\upartial="0\UPM40
      \mathchardef\leqslant="3\AMSa36
      \mathchardef\geqslant="3\AMSa3E
    \fi
  \fi
\fi 

\ifnfsstwo
  \DeclareMathAlphabet{\mathbfit}{OT1}{cmr}{bx}{it}
  \SetMathAlphabet\mathbfit{bold}{OT1}{cmr}{bx}{it}
  \DeclareMathAlphabet{\mathbfss}{OT1}{cmss}{bx}{n}
  \SetMathAlphabet\mathbfss{bold}{OT1}{cmss}{bx}{n}
  \ifAMStwofonts
    \ifCUPmtlplainloaded \else
      \DeclareSymbolFont{UPM}{U}{eur}{m}{n}
      \SetSymbolFont{UPM}{bold}{U}{eur}{b}{n}
      \DeclareSymbolFont{AMSa}{U}{msa}{m}{n}
      \DeclareMathSymbol{\upi}{0}{UPM}{"19}
      \DeclareMathSymbol{\umu}{0}{UPM}{"16}
      \DeclareMathSymbol{\upartial}{0}{UPM}{"40}
      \DeclareMathSymbol{\leqslant}{3}{AMSa}{"36}
      \DeclareMathSymbol{\geqslant}{3}{AMSa}{"3E}
    \fi
  \fi
\fi 

\ifCUPmtlplainloaded \else
  \ifAMStwofonts \else 
    \def\upi{\pi}
    \def\umu{\mu}
    \def\upartial{\partial}
  \fi
\fi

\title{The emission-line pulse pattern in the intermediate polar RX~J0558+53}
\author[Emilios~T.~Harlaftis and Keith Horne]
       {Emilios~T.~Harlaftis,$^1$\thanks{Current address:
       Astronomical Institute, Observatory of Athens, Lofos Koufou, 
       P. Penteli, Athens 152 36, Greece}
       and Keith~Horne$^1$\\
   $^1$School of Physics and Astronomy, University of St. Andrews, St 
       Andrews, KY16 9SS, Scotland, UK \\
      (ehh@astro.noa.gr,       kdh1@st-andrews.ac.uk)\\
       }

\date{Accepted 1998; Received 1997; in original form 1998}
\pagerange{\pageref{firstpage}--\pageref{lastpage}}
\pubyear{1998}

\begin{document}

\maketitle

\label{firstpage}

\begin{abstract} 

We observed the intermediate polar  RX~J0558+53 with the 4.2m WHT  and
find  in the  pulsed emission  lines,  a ``corkscrew'' pattern,  which
indicates a two-pole white  dwarf accretion. The ``corkscrew'' pattern
consists  of two emission-line  pulses,  separated  by half  the white
dwarf  spin  period, and   moving from red   to  blue velocities.  The
detected emission-line  pulsations have an  amplitude of  1.1--2.7 per
cent in the He{\small~II} and Balmer  emission lines on the 545-s spin
period of the  white dwarf which compare to  3.5-4.8 per cent for  the
continuum  double-peak pulsations.  We   image the emission-line pulse
pattern  and is shown  to {\it lag}  the  continuum pulse by 0.12 spin
cycles. We interpret the pattern by invoking an accretion curtain from
the   disrupted, inner disc  to  the two poles  of  the magnetic white
dwarf.  The semi-amplitude of the He{\small~II} pulse of 408$\pm$35 km
s$^{-1}$ can be used to constrain the size of the magnetosphere, $ R
\sim 4.1 \times 10^{4}$ km, and the magnetic moment of the white dwarf
($\sim~2.4 \times  10^{32}$ G~cm$^{3}$).  Power spectra  show dominant
frequencies  at  $2~\omega$  and   $2~(\omega-\Omega)$  which  suggest
reprocessing of the white dwarf's  illuminating beams in the accretion
disc.  Finally,  the  steady He{\small~II}  emission line shows  a
strong sinusoidal component moving  from  red to  blue on  the orbital
period, with a width similar to that expected  from irradiation of the
secondary star. Imaging of the emission lines indicate illuminated locations
at the inner side of the red star and the back side of the accretion disc.

\end{abstract}


\section{Introduction}

Intermediate  polars (IPs) form a  class  of cataclysmic variables and
consist of  a  white-dwarf `pulsar' and a   red dwarf companion.   The
white-dwarf `pulsar' is a magnetic  white dwarf ($10^{4}<B<10^{7}$ G),
with a rotation period between 33 seconds and a fraction of the binary
period, which accretes matter from the red dwarf.  The  way the gas is
directed to the  magnetic poles of the  white dwarf is unclear; either
by the gas stream as in AM Her stars (polars) or via an accretion disc
(Lamb  1988). It is  generally accepted that  for orbital periods $<5$
hours   and  low magnetic-moments,  $\mu   <~10^{33}$  G cm$^{3}$,  an
accretion  disc forms  but  is truncated   at  its inner  edge  by the
magnetic field of  the white dwarf.   Gas from the  inner  edge of the
disc is  then funnelled along magnetic field  lines  onto the poles of
the white  dwarf (for  a   review see  Patterson  1994 and  references
therein).  The exact accretion pattern  is controversial but, assuming
that  the  accretion  is  disc-fed,  several workers   now support the
`accretion curtain' model as the means the disc feeds gas to the white
dwarf (Rosen et al. 1988; Ferrario et al. 1993).

There is   considerable interest in   these  objects because  they are
strong  X-ray  sources,  the  magnetic  field  strongly influences the
accretion flow, and the white dwarf's  spin period can be exploited as
a  diagnostic probe.  In  particular, the X-ray searchlight beams from
the accreting  poles  of the  white  dwarf  can be used  to  probe the
accretion flow.  RX~J0558+53  (V405 Aur) was classified as an
intermediate polar
(V=14.6 mag),  with an orbital  period of  4.15 hours,  soon after its
discovery during the  ROSAT all-sky survey (Haberl  et al. 1994) which
also showed a soft X-ray spectrum more similar to that of
polars  (Haberl and Motch 1995).   Allan et  al. (1996) identified the
spin period of    the white dwarf   (545-s)  from optical  and   X-ray
modulations at 545 and   272 seconds  (see  also Skillman   1996). The
continuum spin-pulse profile consists of two peaks and the periodogram
shows most of the  power on the first harmonic  (272-s) rather than at
the fundamental frequency.

YY Dra (Haswell et  al. 1997) and PQ  Gem  (Hellier et al. 1994)  show
similar  optical double-pulse structure  to RX~J0558+5353.  Weak X-ray
double pulses  are also seen  in GK Per  in  quiescence (Ishida et al.
1992)  and in   XY Ari  (Kamata  and Koyama  1993).   The double-pulse
structure may be explained   by  viewing two accreting   poles through
optically  thin accreting  regions which  show  their largest  optical
depth horizontal   to  the  disc  plane.     Given  the  insight,  the
emission-line  pulsations can provide  in   the accretion pattern,  we
observed RX~J0558+5353 with the  aim to resolve sufficiently  the spin
cycle. Preliminary results  show a double-peak  pulse in He{\small~II}
(Harlaftis \& Horne 1996; Walker et al.  1996 also announced pulses in
the Balmer lines; Still et al. 1998).

\begin{table}
\caption{Emission lines of RX~J0558+5353}
\begin{tabular}{ccrc}
  & Flux & EW & FWHM\\
  & 10$^{-13}$ ergs cm$^{-2}$ s$^{-1}$& \AA & km s$^{-1}$\\
H$\alpha$         &4.13$\pm$0.01 & 604$\pm$1& 743$\pm$12\\
H$\beta$          &2.09$\pm$0.02 & 163$\pm$1& 989$\pm$22\\
H$\gamma$         &1.58$\pm$0.02 & 119$\pm$1& 913$\pm$20\\
He{\small~I} 6678 &0.32$\pm$0.03 &  46$\pm$1& 714$\pm$12\\
He{\small~I} 4713 &0.10$\pm$0.04 &  14$\pm$2& 837$\pm$24\\
He{\small~I} 4471 &0.35$\pm$0.03 &  29$\pm$2& 886$\pm$14\\
He{\small~I} 4388 &0.14$\pm$0.03 &  10$\pm$2& 823$\pm$26\\
He{\small~II} 4686&1.68$\pm$0.02 & 120$\pm$1& 933$\pm$18\\
He{\small~II} 4542&0.15$\pm$0.03 &  13$\pm$2& - \\
He{\small~II} 4200&0.18$\pm$0.04 &   6$\pm$2& 870$\pm$33\\
C{\small~III}/N{\small~III}&0.34$\pm$0.03 & 27$\pm$2 &2094$\pm$66\\
\end{tabular}
\end{table}

\newpage
\begin{figure}
\vspace*{8cm}
\psfig{figure=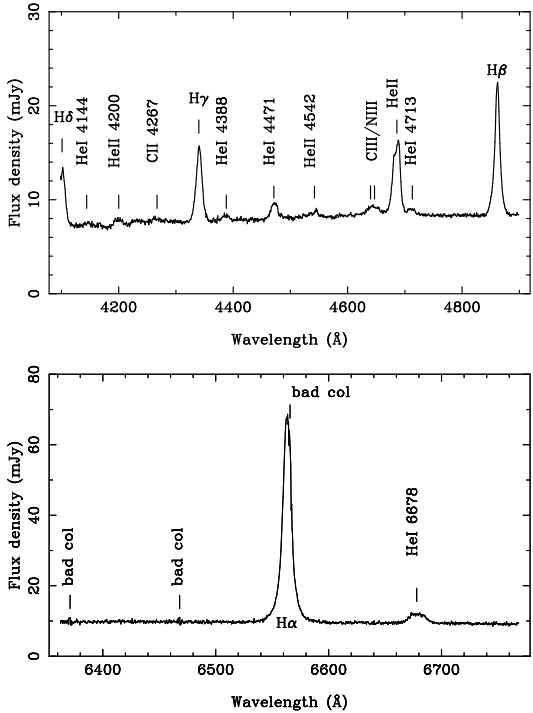,width=8cm}
\caption[]{
The average spectrum of RX~J0558+5353. The most prominent emission lines are 
marked. Parts of the red spectrum affected by 3 bad CCD columns are 
also marked.}
\end{figure}


\newpage

\section{Observations and Data reduction}

We observed RX~J0558+5353 for 3 hours with the WHT 4.2m at La Palma on
17   March 1995 (seeing $\approx~1$ arcsecond).    We used the two-arm
ISIS spectrograph with TEK CCD chips to cover 6360--6770  \AA \ with a
dispersion of   19 km s$^{-1}$  per  pixel and  4100--4900 A  at 52 km
s$^{-1}$ pixel$^{-1}$.  The 30-s exposures were designed to sample the
545-s spin  cycle with 12 spectra. In  total, we obtained 2$\times$251
spectra covering binary   phases  0.2-1.0.  The spectra   were reduced
using  optimal   extraction (Horne       1986) after debiasing     and
flat-fielding  the CCD images.   Sky  subtraction employed polynomials
fitted to sky regions on either side of the  object.  Arc spectra were
extracted from  the same rows as  the object  ones.  The arc lines
drifted by  $<1$~\AA    \ during  the  observations.   The  wavelength
calibration  was performed using  CuNe and CuAr arc  lines for the red
and blue spectra,  respectively,  and is  accurate  to 0.04~\AA.  \  A
comparison star had  been included in  the slit which was subsequently
used to correct  the object spectra for  atmospheric and slit  losses.
The absolute  flux  scale was defined by   observing the flux standard
G191B2B (Oke 1990).

\section{The average Spectrum}

The average spectrum, displayed in Fig.  1, shows a wealth of emission
lines, mainly Balmer,  He{\small~I} and He{\small~II} lines.  Note the
red CCD has 3 bad columns which are marked on the red spectrum in Fig.
1.   The    f$_{\nu}$  continuum   increases   redward,  rising   from
7.15$\pm$0.01 mJy (4130-4180\AA) \ to 8.33$\pm$0.01 mJy (4750-4800\AA)
\ to 9.49$\pm$0.01  mJy (6600-6650\AA).  Table  1 gives the integrated
fluxes ($\pm1000$  km  s$^{-1}$, equivalent widths,   and FWHMs of the
emission lines.  The  Bowen  fluorescence lines at  C{\small~III} 4647
and NIII  4640  together with the  He{\small~II} 4686  line indicate a
high  temperature region  (Schachter  et al.   1991).   In addition to
He{\small~II} 4686,  we  note He{\small~II}  lines from  the Pickering
series (see Fig.  1).  Odd-numbered lines are  seen between the Balmer
lines  whereas    the even-numbered lines   of   the  Pickering series
contribute -- but less than 13 per cent --  to the Balmer-line fluxes.
The Pickering   series of He{\small~II}  are seen  in some AM Her-type
systems (Schachter et al. 1991; see also  the spectrum of the magnetic
nova V1500 Cyg in  Horne \& Schneider  1989).   The flux ratio  of the
Pickering line, He{\small~II} 4542, to  that of He{\small~II} 4686  is
0.10. This is within  the ratio range  0.10-0.19  found in the  AM Her
system E2003+225  (Mukai et al.  1986) but very  different from case B
recombination (0.035;  Seaton  1978). Apparently, the   density of the
emission  region  must play an  important  role  in  this discrepancy.
Still  et al. (1998) obtained  blue spectra of RX~J0558+53 in December
1994   and  red  spectra  in  February  1995.  There  is a  remarkable
difference  between  our spectra  and   those of  Still  et  al.   The
emission-line intensities in  March 1995 have  increased over those in
February 1995 by a factor of 3.3  for the Balmer lines  and by 1.9 for
He{\small~II} (see  Fig. 1 in Still  et al.). The continuum also shows
an increase of 1.9 times in  the blue but only changes  by 15 per cent
in the red.

\begin{figure}
\psfig{figure=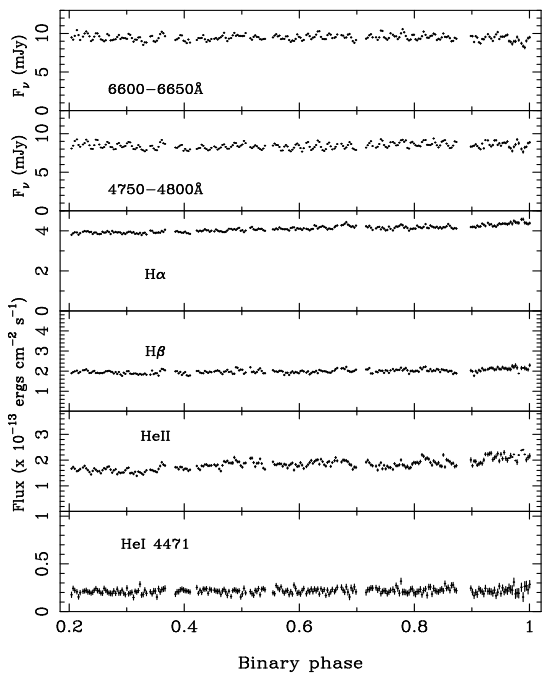,height=10cm,width=8cm}
\caption[]{Orbital variation of continuum and emission lines of the
intermediate polar RX~J0558+53. The line fluxes rise with time whereas 
the continuum does not show any eclipse or orbital-hump variation. 
The 38 low-amplitude pulses visible in the continuum
and partly in He{\small~II} and H$\beta$ are the main features of the 
light curves.}
\end{figure}

\begin{table}
\caption{Pulsation amplitudes}
\begin{tabular}{lr}
band & \% ~~~~\\
0.5-2.0 keV& 12.5$\pm$6.2$^{1}$\\
0.1-0.4 keV& 45.4$\pm$0.4$^{1}$\\
4700-4750\AA& 4.8$\pm$0.2\\
6600-6650\AA& 3.4$\pm$0.2\\
He{\small~II} 4686   &2.7$\pm$0.1\\ 
H$\beta$ &1.8$\pm$0.1\\
H$\gamma$&2.2$\pm$0.1\\
H$\alpha$&1.1$\pm$0.1\\
\end{tabular}

\vspace*{0.5cm}

$^{1}$: Haberl et al. 1994\\
\end{table}

\section{The time series}

The time   variations are shown in    Fig. 2 for  the  continuum bands
6600-6650 \AA  \ and 4750-4800 \AA \  and for the H$\alpha$, H$\beta$,
He{\small~II}, He{\small~I} emission    lines.  Gaps in   these  light
curves   show times when   arc  spectra were obtained.   We  adopt the
following orbital ephemeris (Thorstensen 1997, private communication)

$ T_{o}(HJD) = 2 449474.6446(15) + 0.172624(1)~E$

where  $T_{o}$  is    the    apparent inferior  conjunction   of   the
emission-line  source (blue-to-red  crossing).   A  correction of  0.1
cycles anti-clockwise (see \S~7)  to the apparent inferior conjunction
of the  emission-line source  gives  the  inferior conjunction of  the
secondary star. In general, emission anisotropies on the disc can move
away  the inferior  conjunction of  the emission-line  source from the
inferior conjunction of the secondary star (see,  for example, AE Aqr;
Robinson et al. 1991; Welsh et al.  1993).

There  is  no eclipse   or   any  large  variation  in  the  continuum
light-curves.  However, thirty   eight  pulses  are  visible in    the
continuum  light curves with an  amplitude of 2-6  per cent in the red
and 4-8 per cent in the blue (fitting a sine to all the pulses gives a
mean amplitude  of 3.4  and   4.8 per cent, respectively).   The  blue
pulses  are better  resolved than the  red  (compare, for example, the
second segment of the  data).   In comparison,  Ashoka et al.   (1995)
found a  10.8 per cent  modulation  in white  light. Large variability
between spin cycles  is visible  in many  cases, suggesting that   the
accretion  pattern  may be variable  on   very short time-scales.  The
Balmer and He{\small~II}  lines rise slightly in flux  with time.  The
most   clear pulses  in     the   emission lines   are     visible  in
He{\small~II}.  Most He{\small~II} pulses  are not  in phase with  the
continuum pulses  and some are even in  anti-phase (e.g.  see last two
pulses  in the    second segment).   In    \S~8, we  show   that   the
He{\small~II}  pulses     follow   behind the  continuum     pulses by
0.12$\pm$0.02 cycles.  A few pulses can be traced in the H$\alpha$ and
H$\beta$ light   curves   (see   \S~7  where  the  spin    pulses  are
resolved). The He{\small~I} light curve is also shown though the noise
most likely smears the pulsations out of detection.

\begin{figure}
\psfig{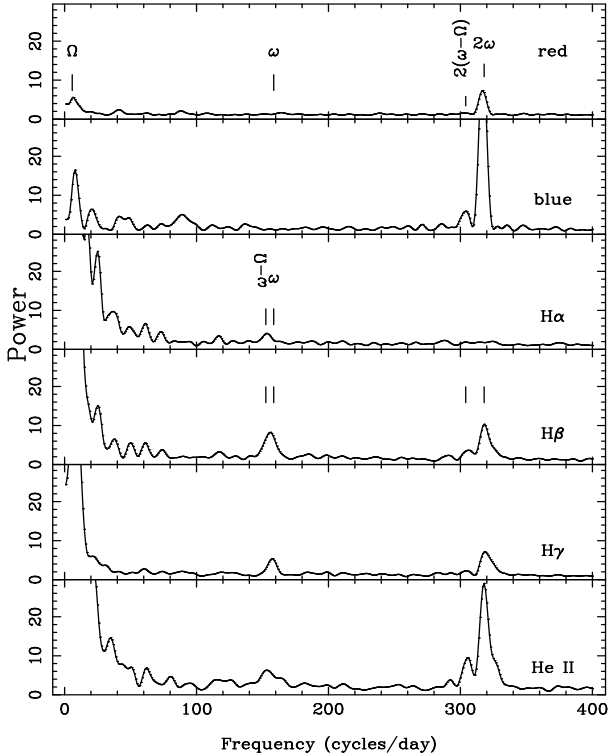}
\caption[]{The Fourier periodogram of the continuum and of the emission lines.
The spin frequency is only evident in H$\beta$ whereas the first harmonic
in dominant in all power spectra except that of H$\alpha$.
An orbital side-band at $2~(\omega-\Omega)$ is also clearly present.
}
\end{figure}

\section{Periodograms: Velocity versus Frequency}

IPs are  characterized by multiple periods as  revealed by analysis of
time  series data. Fourier  analysis  is applied  on the continuum and
emission lines  to reveal the   periodicities of RX~J0558+53  (Scargle
1981).   In Fig. 3  (from top to bottom),  the power spectra are shown
for the red (6600-6650\AA) \ and blue  (4750-4800\AA) \ continuum, and
the three  main    emission  lines ($\pm$1000   km   s$^{-1}$).    The
periodograms are characterized mainly by five peaks at

\begin{itemize}
\item 7.0$\pm$0.6 cycles day$^{-1}$, the dominant peak which
arises from the $\Omega$ orbital frequency (5.8 cycles day$^{-1}$)
and the alias of 7.2 cycles day$^{-1}$ (total observing period)

\item 317.9$\pm0.8$ cycles day$^{-1}$, the second strongest peak
in  all  power  spectra  except   H$\alpha$.  This  is   the 2$\omega$
frequency, where $\omega$ is the  spin frequency, and corresponds to a
period of  271.8$\pm$0.7 seconds, consistent  with the result of Allan
et al. (1996)

\item the side-band frequency at 304.5$\pm$1.0 cycles day$^{-1}$ which is 
clearly  visible in all  periodograms except H$\alpha$. It is strongest
in  He{\small~II}  and we identify  it  with the $2(\omega  - \Omega)$
frequency.

\item 155.7$\pm$1.8 cycles day$^{-1}$ (554.9$\pm$3.9 seconds)  which is 
clearly visible in the H$\beta$,  H$\gamma$ and He{\small~II} emission
lines and is most likely  a blend  of the  $\omega$ frequency (158.5
cycles day$^{-1}$) and the  $\omega -  \Omega$ frequency (152.7  cycles
day$^{-1}$).

\end{itemize}

\begin{figure*}
\psfig{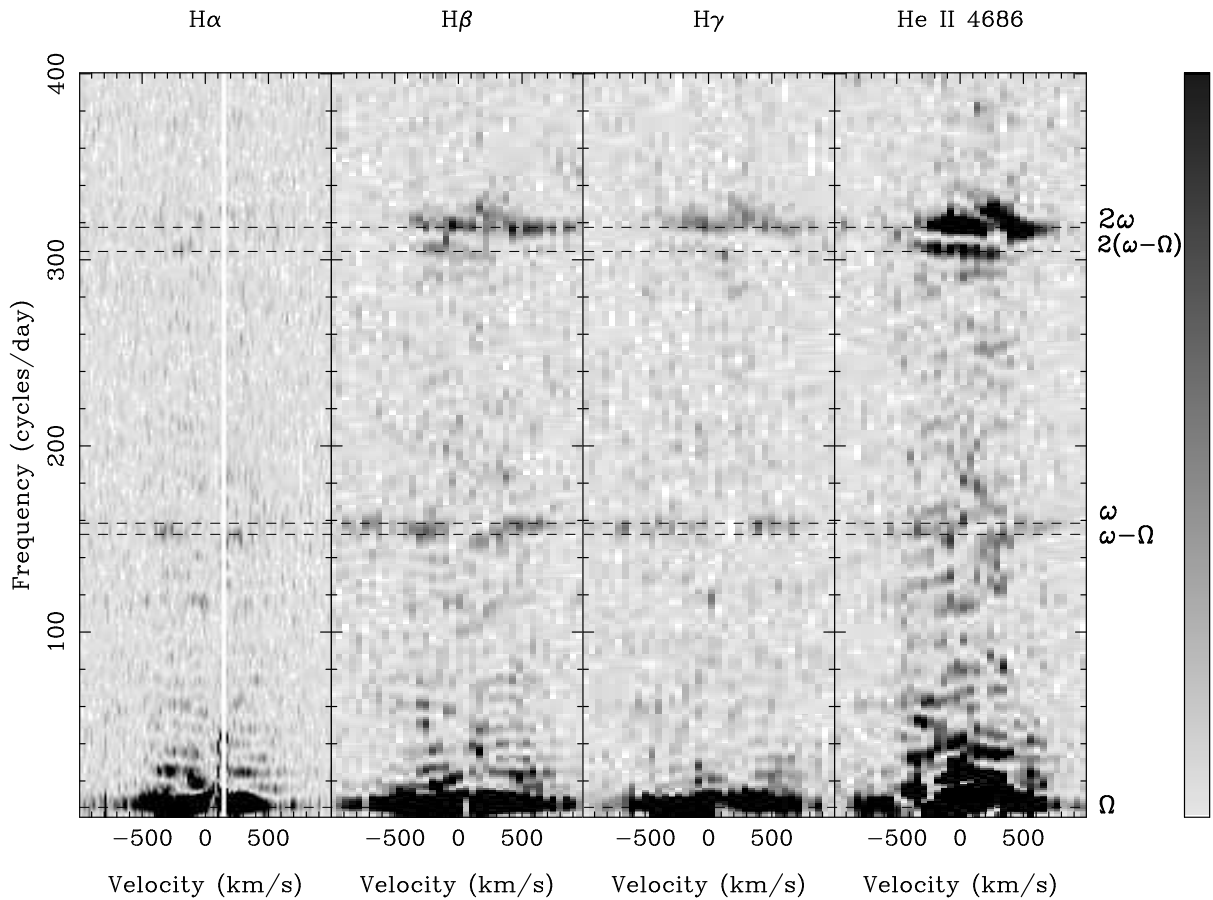}
\caption[]{The Fourier periodogram per velocity bin in the continuum 
and the emission lines. See text.}
\end{figure*}

The above values were extracted  by  fitting Gaussian profiles to  the
individual  periodograms and then averaging  them out. The uncertainty
is indicative of   the  1$\sigma$ scatter of the   individual  values.
Velocity structure  on the emission line profiles  in Fourier space is
resolved  by estimating the power spectrum  per wavelength bin (19 and
52  km   s$^{-1}$ per pixel  for blue    and  red, respectively).  The
resulting   velocity-frequency   diagrams  for   H$\alpha$,  H$\beta$,
H$\gamma$ and He{\small~II} are shown in Fig.  4 (data affected by bad
CCD column  have been subtracted  from  the H$\alpha$  diagram).  Note
that the fourier power periodograms in fig.  3  resulted by adding the
fourier power periodogram for each  pixel of the emission line profile
(or continuum range). Aliases, caused by the  sampling, are evident at
7.2 and 35-40 cycles day$^{-1}$.   Complex structure is evident at the
$2\omega$ frequency  in  H$\beta$,  H$\gamma$ and  He{\small~II}   but
absent in H$\alpha$.  There is a broad component of width FWHM$\approx
380$  km  s$^{-1}$ at --50  km  s${^-1}$.  There  is also a redshifted
component,  centred at 540   km s$^{-1}$, at  the  $2\omega$ frequency
which, below 350 km s$^{-1}$,  shows an increasing frequency from  318
to 330 cycles day$^{-1}$.   The side-band $2~(\omega-\Omega)$ is clear
in He{\small~II} emission, centred at --25 km $^{-1}$  with a width of
FWHM=508 km  s$^{-1}$ (and possible  hints of  emission  in the Balmer
lines).  A     velocity   structure  (5$\sigma$)  at  the    frequency
292.4$\pm$1.2 is probably a sideband frequency to 2$\omega$.  Blue and
red power  components  are visible in  H$\beta$  (5$\sigma$), H$\gamma$
(3$\sigma$) and    possibly  in  H$\alpha$  between   the $\omega$  and
$\omega-\Omega$  frequencies.   The  side-band $\omega-\Omega$  may be
seen at +200-300 km s$^{-1}$  in H$\alpha$, H$\beta$ and He{\small~II}
which coincides  with the approximate velocity  of the secondary star.
A  118 cycles  day$^{-1}$ frequency is  also  visible at the 3$\sigma$
level in all emission lines. We  found no evidence  for the second and
third   harmonics (181.8  and   136.4 seconds).   `Flickering' between
frequencies    0-2$\omega$    is     visible,   particularly        in
He{\small~II}.   However, most of  the   flickering, is attributed  to
aliases of  the sampling (e.g.  $\sim$35-37 cycles/day in H$\beta$ and
He{\small~II}, 23 and 80 cycles/day in He{\small~II}).

\begin{figure*}
\psfig{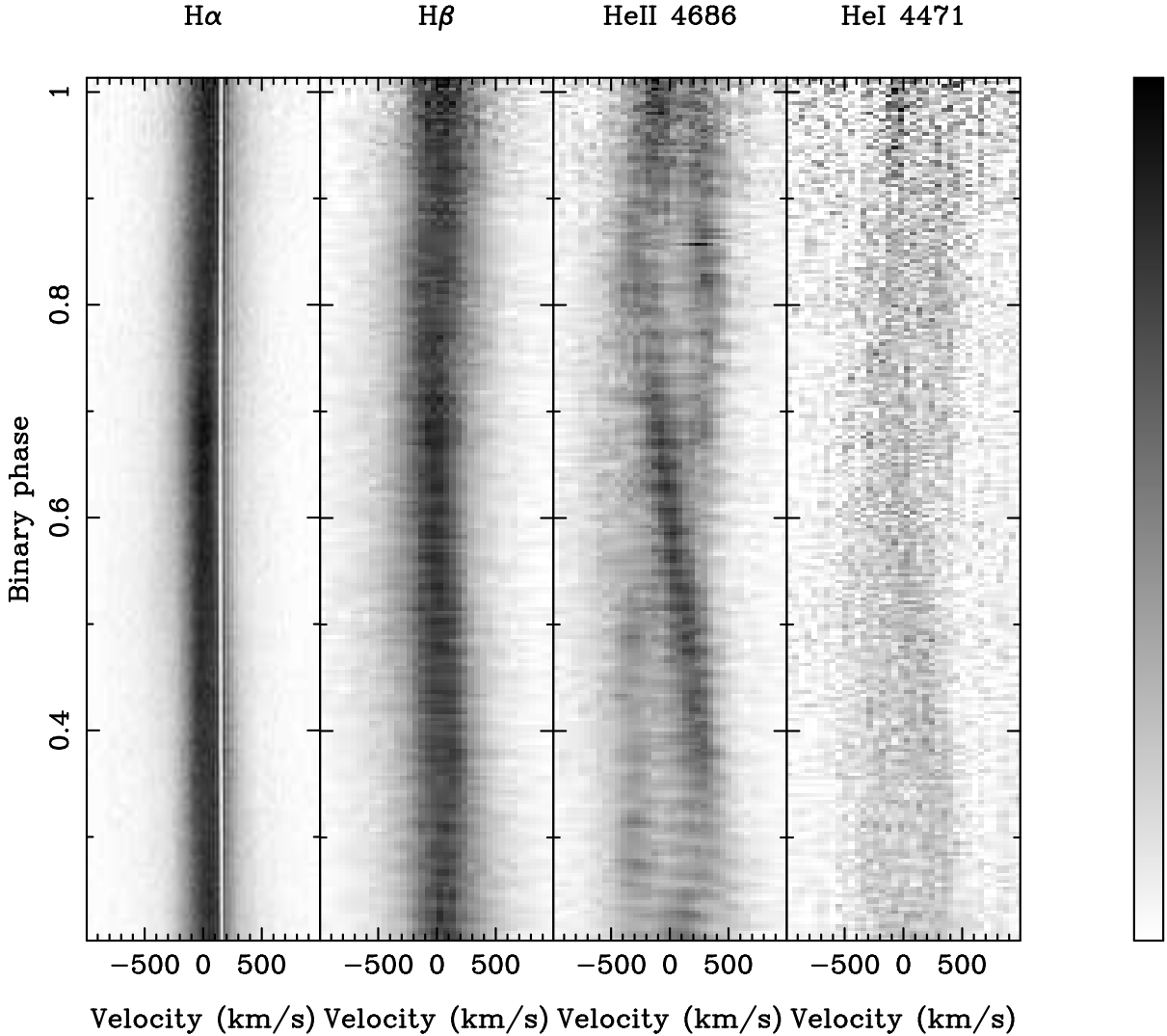}
\caption[]{The trailed profiles of He{\small~I} 4471, He{\small~II} 4686 and
H$\beta$ 17 March 1995. 
The difference between HeI and hydrogen is remarkable.
The sinusoidal He{\small~II} component moving from red to blue is the dominant
feature and can be traced in HeI but not in H$\beta$.
Spin pulses from the white dwarf are also evident in He{\small~II}.
The absolute orbital phase is +0.1 cycles (see text). 
The velocity relative to the line centre along the horizontal
axis is plotted and the orbital phase along the vertical axis.
The intensity scale is adjusted to that of the lines.}
\end{figure*}

\section{The trailed spectra}

The  trailed spectra of the  strongest emission lines are displayed in
Fig.  5.  H$\alpha$  shows a single-peak  profile, in contrast to  its
double-peak  profile only a   month before our  observations (Still et
al. 1997).  Its flux has also increased over that period of time, thus
the double-peak  profile may still be embedded  into the much stronger
single-peak emission component.  The bad  CCD column, that affects the
line profile, can serve as a reference  to demonstrate that the binary
motion in H$\alpha$  is not  clearly visible,  though evident  in  the
Fourier   power spectra.  The  H$\beta$ profiles   show more clearly a
single-peak core    moving with a   low-amplitude   velocity.   Pulses
produced by  the rotating beams of the  white dwarf are visible on the
line profiles.    The  orbital  variation of   the He{\small~II}  4686
profile is complex  and very different from  that of the Balmer lines.
The  average He{\small~II} profile  is a double-peak profile ($\pm$400
km s$^{-1}$).  The trailed spectra then  reveal a prominent sinusoidal
component  which crosses from red to  blue at (relative) phase 0.6 (as
we will see,  absolute phase 0.5)  and shows maximum  strength between
(relative) phases 0.4-0.7.  In addition, another emission component is
either moving in anti-phase to the previous component or is part of an
asymmetric  double-peak  profile.   The He{\small~II} narrow component
shows a broadening of FWHM=230$\pm$28  km s$^{-1}$ which is consistent
with     emission   from the   Roche   lobe  of   the  secondary  star
(Doppler-broadened $\approx200~$   km s$^{-1}$ by the  binary motion).
We identify  this emission with the  inner side of  the red star.  The
above suggests  that the  inferior conjunction  of  the secondary star
(absolute binary phase) leads by 0.1  cycles the emission-line source.
The spin pulses  in the line profiles are  very clear in He{\small~II}
and show the  same repetitive pattern (phase  shift  from red  to blue
velocities;  see  \S~7).    Trace  of the   red-to-blue,   sinusoidal,
He{\small~II} component can also  be discerned in the trailed profiles
of He{\small~I} 4471.

\begin{figure*} 
\psfig{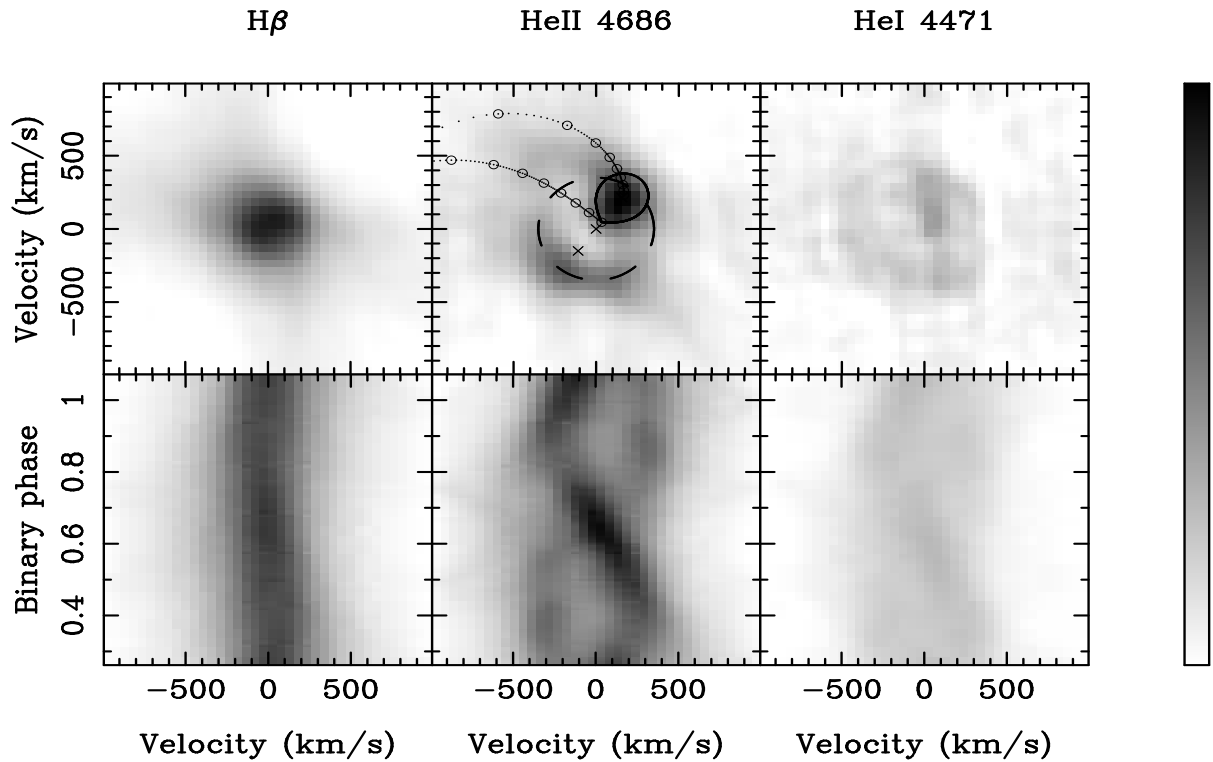}
\caption[]{The top panels display the Doppler maps (filtered back-projection).
The H$\beta$ image displays an almost axi-symmetric distribution centred 
at zero velocity.
The He{\small~II}-image spot, correspoding to the sinusoidal 
emission component moving from red-to-blue,  indicates the 
irradiated  red star. A set of binary parameters has been adopted based 
on this identification ($K_{r}=270$ and $K_{w}=180$ km s$^{-1}$).
A dashed circle at 350 km s$^{-1}$ indicates the velocities of the arc-shaped
emission, moving opposite to the secondary star. 
Steps in units of $R_{L_{1}}$ are marked along the two trajectories 
(ballistic velocities - lower one - and Keplerian velocities along 
gas stream).
The computed data from these images can be compared with the trailed 
spectra of Fig. 3.} 
\end{figure*}

\section{Doppler maps}

We reconstruct images  of  the emission-line distributions using   the
Doppler-shifted line profiles  (trailed  spectra).  The  Doppler image
projected    in  a particular direction   (orbital   phase) is  a line
profile.    The  back-projection   imaging  inverts    the problem and
reconstructs the image from the  line profiles (Horne 1991).  Examples
and techniques of   Doppler    tomography (using either    the  linear
back-projection or maximum  entropy) have been successfully applied in
cataclysmic  variables and X-ray  binaries.  The imaging technique has
resolved emission within the   binary from locations  such as  the red
star  (IP Peg;  Harlaftis  et  al. 1994),   the  gas stream (OY   Car;
Harlaftis  and  Marsh 1996a; see     here, in particular,  for  linear
back-projection image reconstruction),   the  bright spot  (GS2000+25;
Harlaftis et al. 1996b) and culminated recently  with the discovery of
spiral  waves in   the outer accretion  disc  (IP  Peg  during rise to
outburst,  Steeghs  et  al. 1997   IP   Peg during outburst   maximum,
Harlaftis et al. 1998).


The filtered back-projection   images are displayed in   Fig. 6.  High
frequency noise was suppressed by applying a Gaussian smoothing filter
with a FWHM equal to $\approx~90$ km s$^{-1}$.  The H$\beta$ image has
a  nearly axisymmetric  core emission  around zero  velocity (the same
holds  also   for  H$\alpha$).  As  with    the trailed  spectra,  the
He{\small~I} and He{\small~II} images  are similar but differ from the
Balmer-line images.    The helium images   indicate a spot, associated
with  the emission component moving    from red-to-blue, and also   an
emission arc/ring which arises from the other component moving in anti-phase.
The emission spot is most likely  related to the irradiated inner face
of the red  star, which is not  untypical in magnetic CVs (e.g. narrow
component in AM Her stars; Smith 1995).

A  crude velocity estimate of  the He{\small~II} spot from the Doppler
image gives a radial velocity for the inner face of the secondary star
of $\sim250\pm30$ km s$^{-1}$. Using this estimate as a constraint, we
adopt $K_{r}/K_{w}=0.6$ and  $K_{r}+K_{w}=450$ km s$^{-1}$  (otherwise
arbitrary) and a phase offset of 0.1 cycles clockwise. The path of the
gas  stream  and the    Keplerian  velocities along   the   stream are
indicated.  Some emission between the two trajectories is hinted.  The
locations of the stars are also marked (see He{\small~II} image).  The
emission arc/ring  extends   clearly over 0.3  cycles   and is  mainly
confined  between 250--450 km s$^{-1}$  (see circle at 350 km s$^{-1}$
marked on the image). It is located to the back rim  of the disc (i.e.
away from the red star) since it  is related to the emission component
moving in anti-phase to the red star  component.  The bottom panels of
the  figure present the trailed  spectra computed  from the images for
comparison with  the observed  data of Fig.  5.   The Doppler  maps of
Still et al. (1998)  show that the structure,  we reconstruct, is  not
transient  but persists with  time  (i.e. red  star component and back
side of accretion disc).

\begin{figure}
\psfig{figure=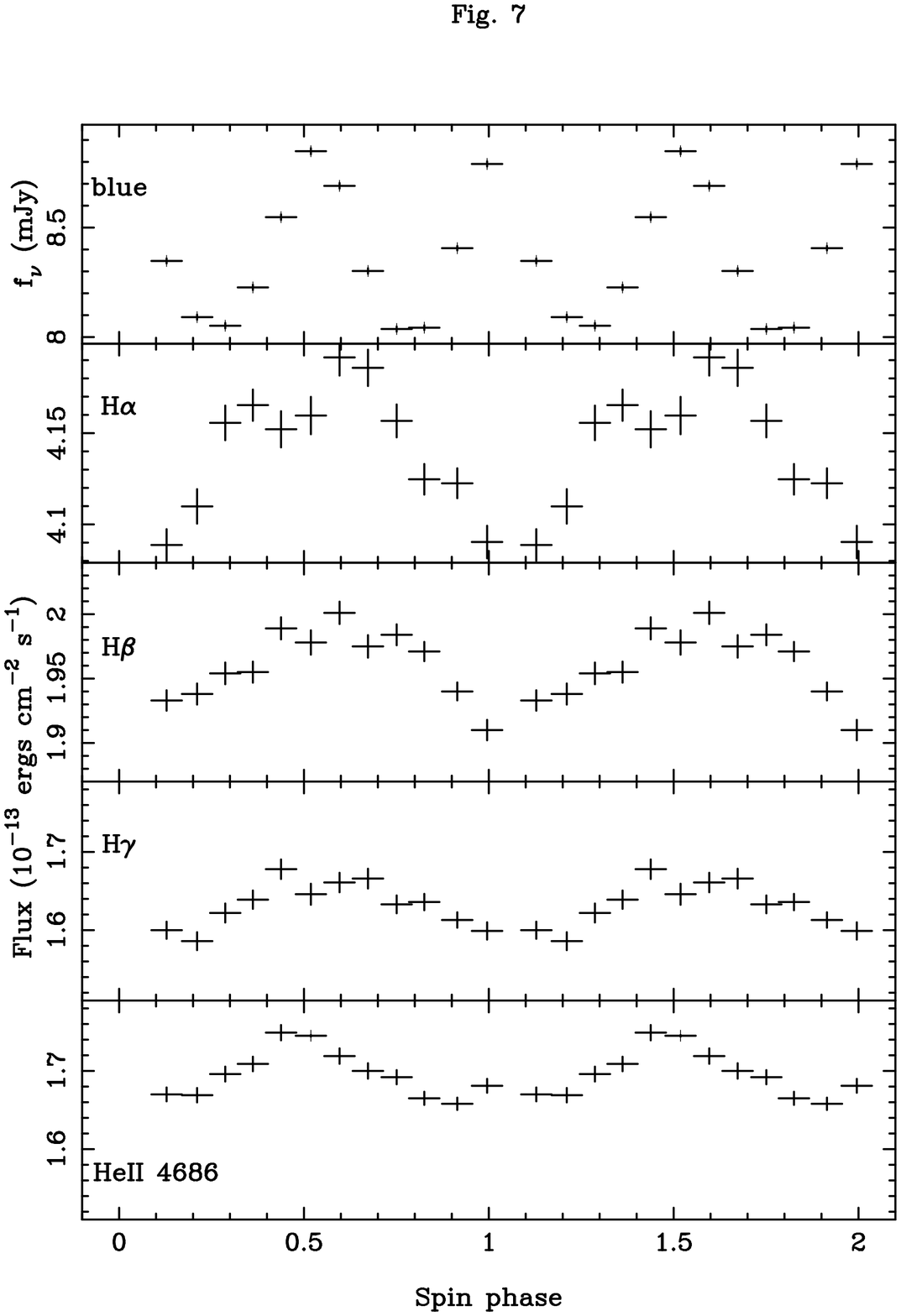,height=10cm}
\caption[]{Spin variation of continuum and emission lines of the
intermediate polar RX~J0558+53. }
\end{figure}

\section{The spin pulses}

Continuum and emission line pulse  profiles folded on the spin  period
of RX~J0558+5353 are presented in Fig. 7.  The 251 spectra were binned
in 12 phases using the spin ephemeris of Allan et al. (1996)

$HJD_{max} = 244 9681.46394(6) + 0.006313143(5) \times E$

(the time resolution is such as to resolve the 545-s spin period in 12
spin phases). The continuum pulse shows two  peaks per cycle at phases
0 and 0.5 (showing the ephemeris is  valid within the errors) and with
an  amplitude of   0.39$\pm$0.01  mJy  (4.7$\pm$0.1  per cent).    The
emission   lines show a   single broad peak  (with apparent structure)
around spin phase 0.5 and a minimum around spin phase 0.0, rather than
a double-peak pulse  structure.   Fitting sines to   the emission-line
variation, we derive amplitudes of 1.1 per cent for H$\alpha$, 1.8 per
cent  for H$\beta$, 2.2  per  cent for H$\gamma$ and   2.7 per cent in
He{\small~II} (with a  typical   error  of 0.1  per cent).     Table 2
presents the pulsation amplitudes for various lines and continuua.

However, the    emission-line pulse structure    is  clarified when we
subtract the  average from  the 12 line   profiles binned on  the spin
period.  The result, displayed in Fig.   8, shows the striking pattern
of two peaks  per cycle crossing the   line profile from red to  blue.
The intensity scale  was set  between -0.5   and 1.5 mJy  in order  to
maximize  the contrast.  Note that the  intensity  scale for all pulse
profile   figures    will be  inversed   (absorption  is  ``dark'') in
comparison to  the binary phase figures  (emission  is ``dark'').  The
double-peak pulse is weakest in H$\alpha$  (the bad CCD column is also
affecting the  trailed spectra) and   strongest in He{\small~II}.  The
two  pulses move from red  to blue velocities  with spin phase, like a
corkscrew, and  their integrated flux  cause the structure  visible at
spin phase  0.5 (see Fig.  7).   For He{\small~II}, the two pulses are
simultaneously visible between 0.5--0.7 spin  phase. There is a slight
asymmetry in He{\small~II} between the two pulse minima (below average
level; dark shade).   The first complete  pulse minimum is longer than
the  second  one in He{\small~II},  a  trend which is  reversed in the
Balmer lines.  Fig.  9 shows the velocities of the He{\small~II} pulse
maxima (`emission'), extracted using Gaussian fits, as they shift with
spin phase.  We derive semi-amplitudes by fitting sinusoids (see Table
3).  The weighted mean semi-amplitude  is $408\pm35$ km s$^{-1}$.  The
He{\small~II} pulse velocities cross zero at spin phases 0.89$\pm$0.02
and 0.26$\pm$0.07 (from  linear fits),  indicating  a lag behind   the
continuum    pulse by a weighted  mean   of 0.12$\pm$0.02 cycles.  The
velocity semi-amplitude  of the pulses  is  408$\pm$35 km s$^{-1}$ and
the FWHM is 318$\pm$17 km s$^{-1}$ (weighted mean; see also Table 3).

We reconstruct  an image of  the  He{\small~II} spin pulses  using the
filtered back-projection technique (with  a Gaussian filter of FWHM=62
km  s$^{-1}$ to suppress high frequency  noise).   Fig. 10 displays an
image   of the   back-projection of   the    pulse data (``corkscrew''
pattern).  The  figure shows  the  velocity distribution of the pulsed
He{\small~II}  emission, as observed from the  white  dwarf, and has a
'quadrapole-like' pattern consisting of maximuma ('bright') and minima
('dark')   at right angles.   The phase  direction  of  the minima and
maxima of the  two pulses is marked in   order to aid the  reader. The
first pulse  minimum and maximum  in Fig. 9 is  represented by the two
right velocity quadrants in Fig. 10.  In addition, this image gives an
alternative  method of  confirming our   previous measurements.    The
direction  of the two pulse maxima  is along the   axis defined by the
spin  phases $\approx$0.38 and 0.88,  confirming   the 0.12 phase  lag
found with respect to the continuum pulses.

\begin{figure*}
\psfig{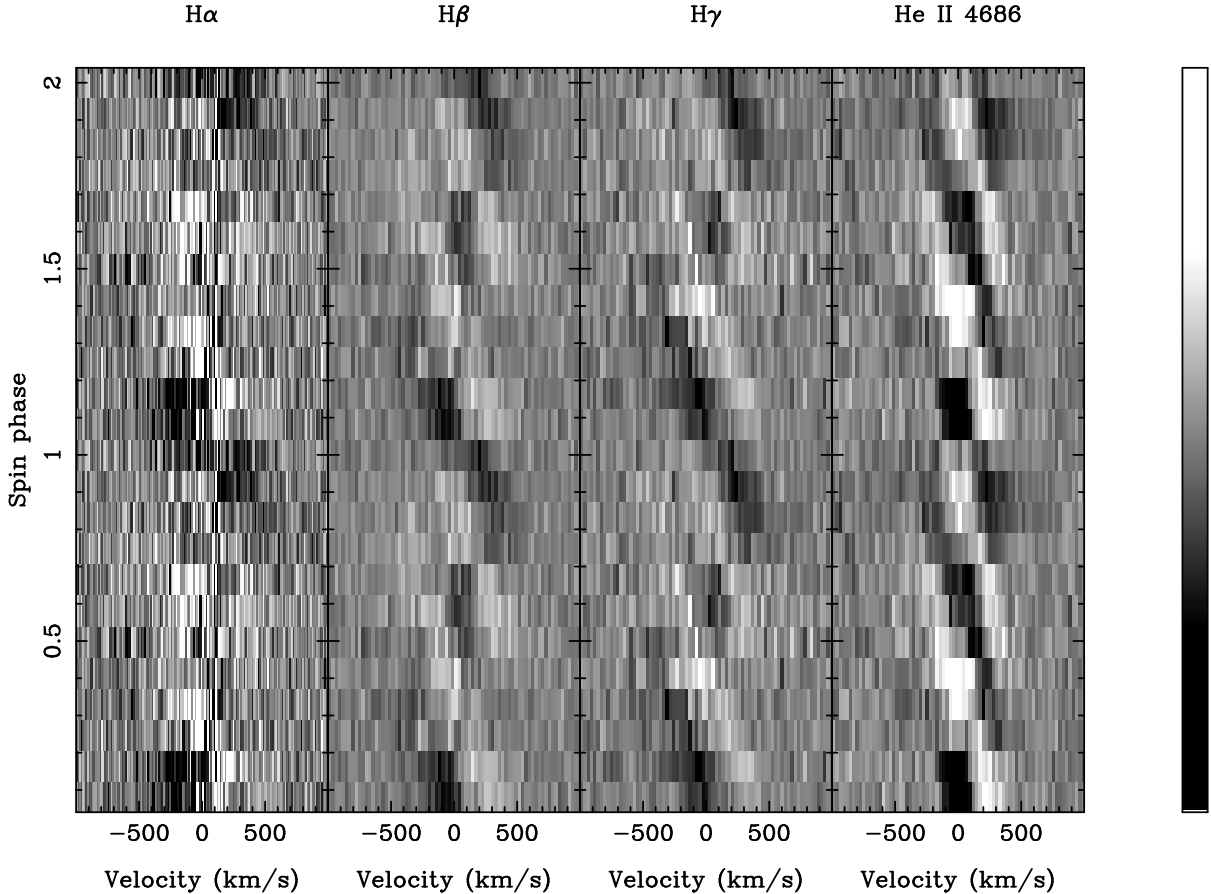}
\caption[]{Spin pulses for the main emission lines.
A double-peak pulse is evident, strongest in He{\small~II}.}
\end{figure*}

\section{Discussion}

\subsection{Beam illumination of the accretion disc, gas stream and red star}

The frequencies detected   in the periodograms  probe the illuminating
effects of the white dwarf's beams into the  binary's components.  For
example, lack of the $2\omega - \Omega$ peak suggests that there is an
accretion disc (Hellier 1992; Wynn \& King 1992).  Further, prevalence
of the  $2~(\omega-~\Omega)$ and $2~\omega$ frequencies  indicate both
disc-  and stream-fed   emission from  two diametrically-opposed poles
with  similar emission     properties (Norton 1996).     Norton (1996)
presents  a  simple  model  for the  X-ray  emission based   on cosine
emission and attenuation components from which he constructs the X-ray
light  curve  and power   spectrum (adopting the   optical light-curve
analysis by  Warner  1986).  Different combinations of  frequencies is
then suggestive of specific illuminated  patterns.  For example, 
a simple, disc-fed emission has most power in the $2\omega$ frequency.

The dominance  of the  frequencies $2~(\omega-~\Omega)$ and $2~\omega$
(away from the $\Omega$ frequency  region) suggests that the side-band
$2~(\omega-\Omega)$ frequency arises  mainly from the beat between the
white  dwarf rotation  and the  orbital  modulation of the gas stream,
which probably encounters the magnetosphere through disc overflow (see
King  \& Lasota  1991 for disc  overflow).   Note that  there is power
spread in He{\small~II}  up to the  $2~\omega$ frequency which implies
that emission arises from a wide range of periodicities independent of
velocity and may represent beam reprocessing in  the accretion disc and/or
the region close  to the white dwarf.   The  broad He{\small~II} power
component  at the $2~\omega$ frequency,  centred at  --50 km s${^-1}$,
may well  arise on  the  white dwarf.  The  peculiar red-shifted power
structure shows a frequency increase (up to 330 cycles day$^{-1}$ with
decreasing velocity below  350 km s$^{-1}$, which  is seen in H$\beta$
and H$\gamma$ as well.  Red-shifted  components are expected from such
a system  either from the red star  at 250$\pm$30  km s$^{-1}$ or from
the inwards  gas flow  to the  white dwarf   along the magnetic  field
lines.   Any reprocessing signal as seen from the secondary star
will  appear  at   a  frequency $\omega_{irr}=\omega-\Omega\approx153$
cycles/day.  This frequency  may be present,  redshifted by 300$\pm$50
km s$^{-1}$ in  He{\small~II} and possibly  in  H$\beta$ as well  (see
Fig. 4).   

Indeed,   we  have identified   emission   from the  red  star  on the
He{\small~II} Doppler image.  The    emission arc, extending   between
140$^{o}$--230$^{o}$, and  lying around 350 km  s$^{-1}$, is then part
of the accretion disc behind the white dwarf. This may partly coincide
with the  location    of  reprocessing evidenced by     the  side-band
$2(\omega-\Omega)$.  Such a location could be the direct impact of the
gas  stream on the magnetosphere/inner disc  as envisaged through disc
overflow (Lubow 1989;   Hellier 1993).  Lubow  finds   that the impact
location is  always  between 140$^{o}$-150$^{o}$, independent  of mass
ratios, and lies at a distance  of $6.86~R_{wd}~M_{0.6}^{-0.27}$ for a
mass ratio of $q=0.5$  ($R_{wd}$ is the radius  of the white dwarf and
$M_{0.6}$ is the mass of the  white dwarf in units of 0.6~$M_{\odot}$.
The    free-fall    velocity at  such   a   distance   is $V~\sim 1700
~M_{0.6}^{0.8} $    km s$^{-1}$. Based on   the  above discussion, the
arc-like component on the Doppler image may well represent the shocked
region  (velocities drop from  1700 to 350  km s$^{-1}$) of the stream
overflow onto  the  transition region of the   disc (Keplerian flow to
angular velocity of the white dwarf).

\begin{table}
\caption{He~II pulse properties$^{*}$}
\begin{tabular}{ccccc}
pulse & & FWHM & K &  $\phi$ \\
      & & km s$^{-1}$& km s$^{-1}$& \\
first &maximum&310$\pm$47  & 413$\pm$41 &  0.89$\pm$0.02 \\
      &minimum&319$\pm$181 & 349$\pm$71 &  0.56$\pm$0.04 \\
second&maximum&327$\pm$60  & 339$\pm$68 &  0.26$\pm$0.07 \\
      &minimum&411$\pm$172 & 448$\pm$41 &  0.11$\pm$0.03 \\
\end{tabular}

$^{*}$ : extracted from sinusoidal fits to the data

\end{table}

\begin{figure}
\psfig{figure=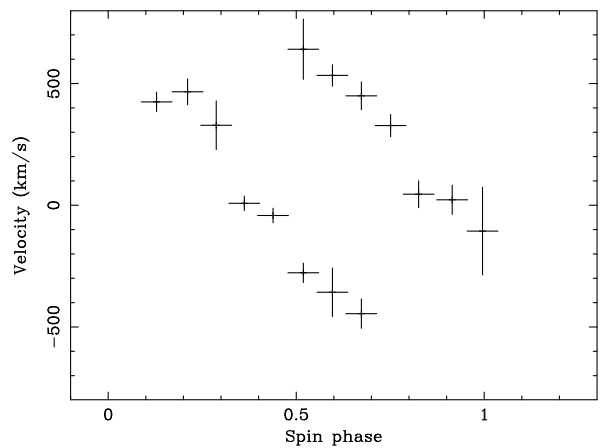,width=10cm,angle=-90}
\caption[]{The velocities of the two He{\small~II} pulses with spin phase.}
\end{figure}

\subsection{The ``corkscrew'' pattern : A two-pole accretion}

A two-pole accretion model, based on the double-peak pulse profile and
the dominance of the  first harmonic in  the  optical and X-ray  power
spectra, has been proposed for RX~J0558+5353 (Allan et al. 1996).  Our
data show pulses (Fig. 2) which at times are different between odd and
even  ones.  Together  with the   observation that  the  power of  the
emission-line pulses  appears    mainly at the  first  harmonic, our
spectra corroborate for a  two-pole  accretion region  as well.    The
He{\small~II} pulse shows a semi-amplitude  of 408$\pm$35 km s$^{-1}$.
For a   Keplerian velocity of  408  km s$^{-1}$,  the location  of the
He{\small~II} pulsations would be placed   beyond the tidal radius  of
the accretion disc.  In addition, such a location for a highly-ionized
region is problematic. He{\small~II}  pulsations are generally related
to  the transition region between the  disc and  the magnetic field of
the white dwarf (Lamb  1988), and in particular  the accretion-curtain
model seems plausible  to interpret the  double-pulse of He{\small~II}
(EX Hya in Rosen et al. 1988; AO Psc in Hellier  et al. 1991; see also
Fig. 11).

Each   pulse maximum   in the   continuum,    most likely,  represents
illumination of gas  by the X-ray rotating beams  in a region  between
the   accretion column  and   the  disc's  transition  region.    Each
He{\small~II}   pulse   may then represent    reprocessed gas slightly
further out from the continuum region which  is supported by the phase
delay of 0.1  spin  cycles. For zero  phase  delay, the  continuum and
emission regions would  lie in the  same axis  (radial infall).  For a
phase delay  of 0.25 cycles,  the corotation  velocity of the emission
region would be  perpendicular to the  radial infall of the  continuum
region. Therefore, the He{\small~II} emission  region lies in a region
between  corotation and radial infall,  and  most likely, tracing  the
magnetic field lines.  Such a delay is consistent with the field lines
being swept back by the perturbed flow of disc matter.

The nearly symmetric pulses in  the continuum indicate accreting poles
of equal brightness  at  anti-diametric locations.  The  He{\small~II}
pulses  are very asymmetric with  the first  pulse dominating over the
second one (flux is higher by a factor of 2).  The width of the pulses
is identical, 318$\pm$17 km s$^{-1}$, which indicates a common origin,
and   sets  upper  limits    for  the   broadening mechanism    in the
He{\small~II} emission-region, perhaps the scale of turbulence (change
from  circular motion  to quasi-radial)  or  the sound velocity ($T  <
10^{6}$~K). While one pulse  is in maximum  blueshift, the other is in
maximum  redshift (``corkscrew'') 
which  is  entirely  consistent  with  an  accretion
curtain.  The maximum  of the first pulse  is viewed directly from the
(lower) pole  whilst   the curtain  blocks   any  visibility   of  the
He{\small~II} region from the upper pole,  resulting in the minimum of
the second  pulse  (in the average-subtracted  line  profiles).   This
varying  view  of the  two  inner parts of   the  curtain explains the
quadrapole-like pattern  of the pulsed  emission in  Fig. 10.  Partial
overlap  of the pulses between  spin phases 0.5--0.7 suggests that the
azimuthal extent  is large enough so  that the two He{\small~II} pulse
locations  can be viewed simultaneously at  times. The  phase range of
the pulses  also suggests that the azimuthal  extent of  the accretion
curtain in  the  continuum is smaller than  that  of the He{\small~II}
emission regions (Fig. 7).

\begin{figure}
\psfig{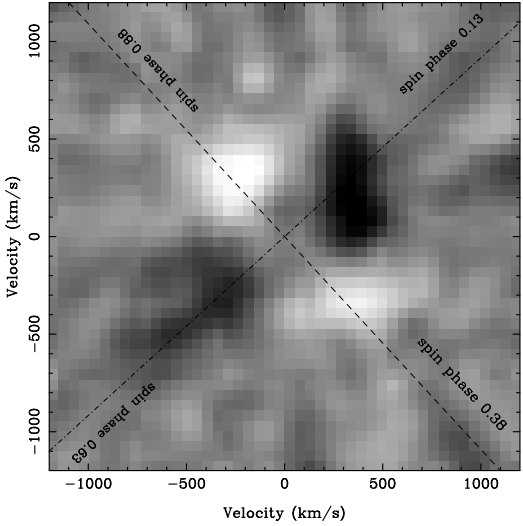}
\caption[]{The Doppler map of the pulsed  He{\small~II} emission,
as  viewed from  the white  dwarf (0,0).  The  back  projection of the
He{\small~II}  pulsed emission  component  produces a  quadrapole-like
velocity distribution, consisting of the minima ('dark' shade) and the
maxima ('bright' shade) of  the two pulses. The  spin phases where the
above are more pronounced are also  marked, for clarity and comparison
with the trailed spectra in Fig. 8.
The emission line pulse lags behind the continuum pulse by 0.12 spin cycles.}
\end{figure}

\subsection{The magnetic moment of the white dwarf}

We associate the emission-line pulsations with material which has lost
its Keplerian  flow due to  the magnetic field's  coupling and is most
likely close to  radial free-fall along the  magnetic field lines (see
previous discussion   about  phase  delay).  Then,  the  He{\small~II}
emission region will be located within  the magnetosphere of the white
dwarf and will rotate with the spin period of 545 seconds.  This gives

$ V = V_{obs}/~\sin~i = \frac{2~\pi~R}{P} $ km s$^{-1}$,

where  $V_{obs}  = 408$  km s$^{-1}$, $P  =545$  seconds  is the spin
period.  There  is no eclipse in the  X-ray and optical  light curves.
On the  other  hand,  the  two accreting  poles  of  the   white dwarf
(double-peak  pulse) can be   viewed only in high-inclination  systems
($50^{o}-70^{o}$) and this is  most likely the case for RX~J0558+5353.
The pulsations become visible only  for favourable orientations of the
rotating searchlight  beams    to the line-of-sight.  Truly,   this is
supported by   a correlation between  inclination and  pulse amplitude
(Hellier  and  Mason   1990).  For a   free-fall   velocity of  408 km
s$^{-1}$/ $\sin~60^{o} = 471$ km s$^{-1}$, the distance is

$ R \sim 4.1 \times 10^{4}$ km or $ R \sim 4.9~R_{wd}~M^{+0.6}_{0.6} $

from the white  dwarf, assuming an inclination  of 60$^{o}$ and  using
$R_{wd}  \sim   0.84  \times 10^{4}~M^{-0.6}_{0.6}  $   km  (Hamada \&
Salpeter  1961).   This may  well represent  the magnetospheric radius
where  the  disc is  disrupted,  $R=R_{mag}$,  half the Alfven radius,
which gives

$ R_{mag} \sim 0.26 ~R_{wd}~ ( M_{0.6}^{0.91}~~L_{33}^{-2/7}~
\mu_{30}^{4/7}) = 4.1 \times 10^{4}$ km

where $L_{33}$ is the  luminosity in units  of $10^{33}$ ergs s$^{-1}$
and $\mu_{30}$ is the magnetic moment in units of $10^{30}$ G cm$^{3}$
(Ghosh and  Lamb  1978).  The  impact  of the  stream overflow on  the
disc's    transition      region  is   slightly     further   out   at
$6.86~R_{wd}~M_{0.6}^{-0.27}$ ($q=0.5$) and  may have lower velocities
of  $\sim$350 km s$^{-1}$.  The  corotation radius  $R_{co}$ (Frank et
al. 1992), $R_{co} \sim 10~R_{wd}~M^{-0.27}_{0.6}$,
is larger than the magnetosphere (slow rotator) for accretion to occur.

Haberl      et   al.  (1994)    derive  an     observed  luminosity of
$0.18~\times~10^{32}~ (\frac{d}{300~{\rm pc}})^{2}$ ergs s$^{-1}$ from
the X-ray flux between 0.1-2.4 keV.  Fitting the X-ray spectrum with a
blackbody of 57  eV (soft component)  and a bremsstrahlung spectrum of
10   keV     (hard       component)  gives     a      luminosity    of
$1.7~\times~10^{33}~(\frac{d}{300~{\rm pc}})^{2}$ ergs s$^{-1}$ for an
absorption column density  of $6~\times~~10^{20}$  H~cm$^{-2}$ (Haberl
et al. 1994). This can be used to estimate the  magnetic moment of the
white  dwarf of $2.4~\times~10^{32}$~G~cm$^{3}$  (or $B \sim 0.5$~MG),
assuming a $M_{0.6}  = 1$, $L_{33}=1$, $d =  300$ pc and a  negligible
soft X-ray  luminosity compared to the  hard X-ray component.   Such a
magnetic moment is not inconsistent with the existence of an accretion
disc, evidence for  which we found  from the periodogram analysis (IPs
with $\mu  > 10^{33}$  G cm$^{3}$  and orbital  periods of $<~5$ hours
cannot form discs; Hameury  et al.  1986).   On the other hand, Haberl
and Motch (1995) suggested that  RX~J0558+5353 resembles AM Her  stars
from its  strong  soft X-ray component  and  may thus provide  a  link
between the  high  magnetic  field  intermediate polars  (BG   CMi and
PQ Gem; Penning   et   al. 1984   and Piirola  et    al. 1993)  and
low-magnetic field AM  Her  stars ($<10$  MG; Schwope 1995).   This is
further strengthened by the   presence of the He{\small~II}  Pickering
series in the  blue spectrum.  Thus, the magnetic  field of  the white
dwarf is  very likely   detectable and  should   be aimed with  future
observations.

\begin{figure}
\psfig{figure=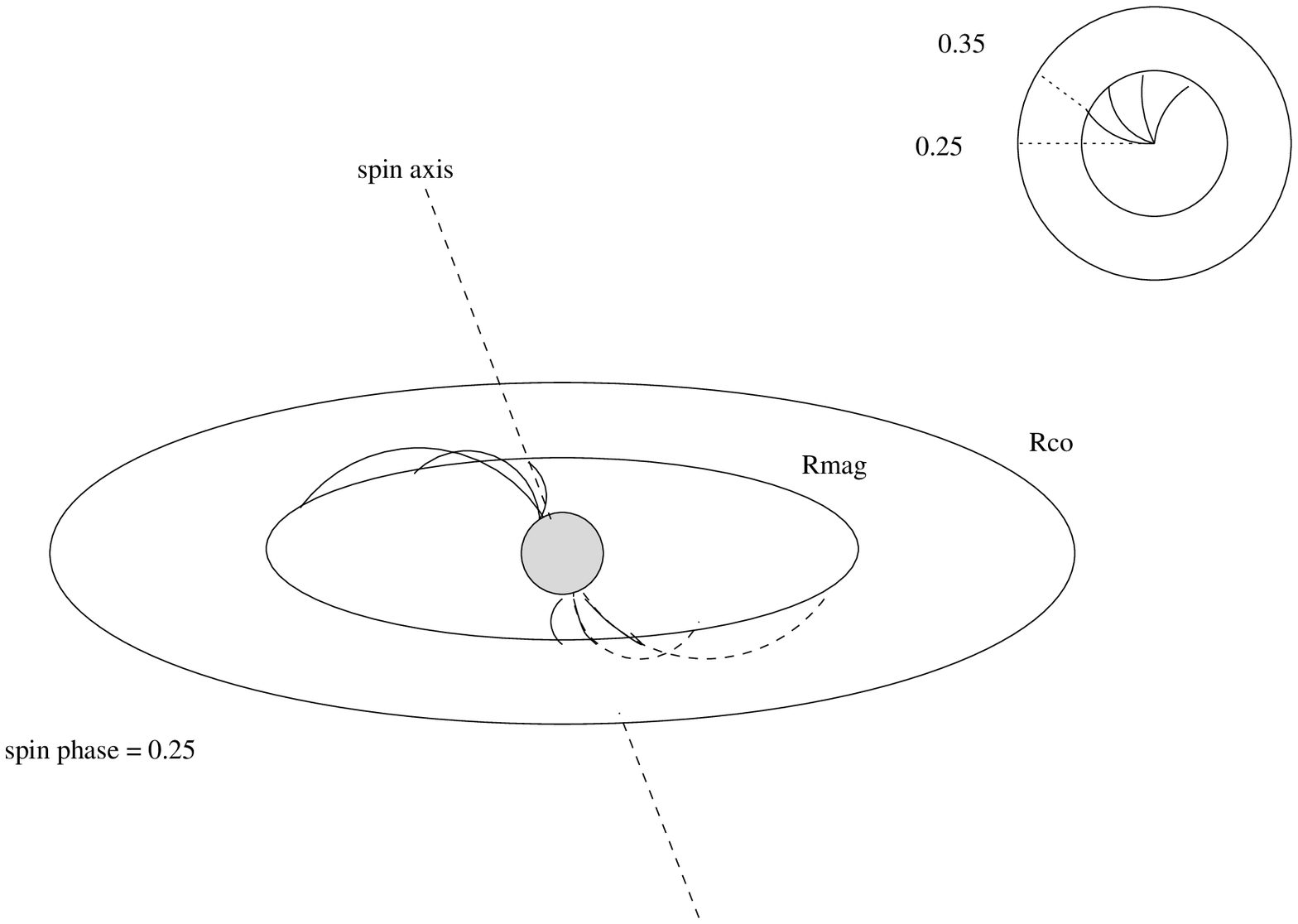,width=8cm,angle=-90}
\caption[]{A schematic diagram of the possible accretion geometry
of RX~J0558+5353. The transition region defined by the corotation radius,
$R_{co}$ is outlined  ($\sim10~R_{wd}$). The inner radius is the
magnetospheric radius, $R_{mag}$, at $\sim~4.9~R_{wd}$, where the 
accretion 
curtain starts. A top view is also shown at the inset to show the
phase delay between the continuum and He{\small~II} regions.}
\end{figure}

\section{Conclusions}

Our observations of the new intermediate polar RX~J0558+5353 reveal

\begin{itemize}
\item a strong, single-peak H$\alpha$ profile in contrast 
to a much weaker, double-peak profile one month before our observations,
indicating a transient H$\alpha$ emission, filling in the double-peak profile
\item a strong He{\small~II} emission profile which shows two emission
components; one at a semi-amplitude of 250$\pm$30 km s$^{-1}$
moving from red-to-blue velocities and with a maximum line strength 
at zero velocity (irradiated secondary star) and another, 
moving in anti-phase,
with velocities of 350$\pm$100 km s$^{-1}$ and related 
to the back side of the accretion disc
\item in addition to the steady line emission, there are
double-peak spin pulsation amplitudes of 1.1 per cent in H$\alpha$, 
1.8 per cent in H$\beta$, 2.2 per cent in H$\gamma$ and 
2.7 per cent in He{\small~II}
\item the dominant frequencies in the spectra are at
$2\omega$ and the side-band $2\omega-2\Omega$ in both
the continuum and the emission lines
\item the emission-line pulsations lag by 0.12$\pm$0.02 in spin phase
relative to the continuum pulsations
\item the nearly symmetric pulses in the continuum become
very asymmetric in the emission lines with the first pulse dominating
over the second one (while they overlap between spin phases 0.5--0.7).
\item The two pulses show a distinctive pattern, the ``corkscrew''
pattern, moving  from  red   to  blue,  twice  per  spin   cycle.  The
velocity-phase shift of the  He{\small~II} pulses has a semi-amplitude
of $\pm408\pm35$  km  s$^{-1}$ and a  FWHM of  318$\pm$17 km s$^{-1}$.
The former gives  an estimate of  the  size of the  magnetosphere, $ R
\sim 4.1 \times 10^{4}$ km, and the magnetic moment of the white dwarf
of $2.4~\times~10^{32}$~G~cm$^{3}$.

\end{itemize}
 
The system's  inclination, the angle  between the orbital and magnetic
axes and asymmetries in  the accretion flow,  such as azimuthal extent
and optical depth  variations as viewed from  different angles, can be
combined  to model  the observed He{\small~II}  pulses. In particular,
the velocity-frequency periodograms provide a  powerful probe into the
accretion pattern which still has to be understood and modelled.

\section*{Acknowledgments}
The observations were taken under the SERVICE programme on the William
Herschel Telescope which is operated by the Royal Observatories of the
United Kingdom at the Spanish Observatorio del  Roque de los Muchachos
of  the Institute de Astrofisica  de Canarias.  The data reduction and
analysis was carried  out at  the  St. Andrews  STARLINK node. Use  of
MOLLY, an  interactive binary analysis   package, developed largely by
T. Marsh, is acknowledged.

\label{lastpage}

\end{document}